\documentclass[journal=jacsat,manuscript=article]{achemso}

\usepackage[version=3]{mhchem} % Formula subscripts using \ce{}

\usepackage{balance}
\usepackage{mathptmx}
\usepackage{amsmath}
\usepackage{sectsty}
\usepackage{graphicx} 
\usepackage{lastpage}
\usepackage{dcolumn}% Align table columns on decimal point
\usepackage{bm}% bold math
\usepackage{float}
\usepackage{fancyhdr}
\usepackage{fnpos}

\author{Jincheng Yue}
\affiliation{Institute of High Pressure Physics, School of Physical Science and Technology, Ningbo University, Ningbo, 315211, China}
\alsoaffiliation{These authors contributed equally}

\author{Jiongzhi Zheng}
\affiliation{Thayer School of Engineering, Dartmouth College, Hanover, New Hampshire, 03755, USA}
\alsoaffiliation[Second University]
{Department of Mechanical and Aerospace Engineering, The Hong Kong University of Science and Technology, Clear Water Bay, Kowloon, 999077, Hong Kong}
\alsoaffiliation{These authors contributed equally}
\email{jiongzhi.zheng@dartmouth.edu}

\author{Junda Li}
\affiliation{Institute of High Pressure Physics, School of Physical Science and Technology, Ningbo University, Ningbo, 315211, China}

\author{Siqi Guo}
\affiliation{Institute of High Pressure Physics, School of Physical Science and Technology, Ningbo University, Ningbo, 315211, China}

\author{Wenling Ren}
\affiliation{Institute of Materials Science, Technical University of Darmstadt, Darmstadt, 64287, Germany}

\author{Han Liu}
\affiliation{School of Control Science and Engineering, Dalian University of Technology, Dalian, 116024, China}

\author{Yanhui Liu}
\affiliation
{Institute of High Pressure Physics, School of Physical Science and Technology, Ningbo University, Ningbo, 315211, China}
\email{liuyanhui@nbu.edu.cn}

\author{Tian Cui}
\affiliation
{Institute of High Pressure Physics, School of Physical Science and Technology, Ningbo University, Ningbo,  315211, China}
\alsoaffiliation[Second University]
{State Key Laboratory of Superhard Materials, College of Physics, Jilin University, Changchun, 130012, China}
\email{cuitian@nbu.edu.cn}

\title[An \textsf{achemso} demo]
  {\textbf{Ultra-low glassy thermal conductivity and controllable, promising thermoelectric properties in crystalline \textit{o}-CsCu$_5$S$_3$}}

\begin{document}
\clearpage
\begin{abstract}
\par We thoroughly investigate the lattice anharmonic dynamics and microscopic mechanisms of the thermal and electronic transport characteristics in orthorhombic \textit{o}-CsCu$_5$S$_3$ at the atomic level. Taking into account the phonon energy shifts and the wave-like tunneling phonon channel, we predict an ultra-low $\kappa_\mathrm{L}$ of ~0.42 w/mK at 300 K with an extremely weak temperature dependence following $\sim T^{-0.33}$. These findings agree well with experimental values along with the parallel to the Bridgman growth direction. The $\kappa_\mathrm{L}$ in \textit{o}-CsCu$_5$S$_3$ is suppressed down to the amorphous limit, primarily due to the unconventional Cu–S bonding induced by the $p$-$d$ hybridization antibonding state, coupled with the stochastic oscillation of Cs atoms. The nonstandard temperature dependence of $\kappa_\mathrm{L}$ can be traced back to the critical or dominant role of wave-like tunneling of phonon contributions in thermal transport. Moreover, the $p$-$d$ hybridization of Cu(3)–S bonding results in the formation of a valence band with ‘pudding-mold’ and high-degeneracy valleys, ensuring highly efficient electron transport characteristics. By properly adjusting the carrier concentration, excellent thermoelectric performance is achieved, with a maximum thermoelectric conversion efficiency of 18.4$\%$ observed at 800 K in \textit{p}-type \textit{o}-CsCu$_5$S$_3$. Our work not only elucidates the anomalous electronic and thermal transport behavior in the copper-based chalcogenide \textit{o}-CsCu$_5$S$_3$ but also provides insights for manipulating its thermal and electronic properties for potential thermoelectric applications.
\end{abstract}
\clearpage
\section{\textbf{INTRODUCTION}}
\par Thermoelectric materials, directly converting heat into electricity, are presently being explored for various feasible waste heat recovery systems, including space-based applications and thermal data storage devices~\cite{massetti2021unconventional, russ2016organic}. Generally, the thermoelectric maximum efficiency of thermoelectric materials is characterized by the dimensionless figure of merit, \textit{ZT}=\textit{S}$^2$$\sigma$/$\kappa$, where \textit{S}, $\sigma$, and $\kappa$ represent the Seebeck coefficient, electrical conductivity, and thermal conductivity, respectively.
A central focus in thermoelectric materials research is to notably enhance conversion efficiency by minimizing irreversible heat transport and preserving favorable electrical transport properties~\cite{yue2023pressure, yue2023significantly}. Thus, the proposal of the phonon glass electron-crystal (PGEC) concept aims to identify high-performance thermoelectric materials~\cite{beekman2015better}. In this paradigm, ordered crystals preserve favorable electronic properties while demonstrating lattice thermal conductivity ($\kappa_\mathrm{L}$) comparable to amorphous solids or glasses. Nonetheless, the pursuit of a net increase in the power factor poses a challenge to enhancing thermoelectric performance due to the strong coupling among electrical transport~\cite{mi2015enhancing, guo2024novel}. Therefore, a comprehensive exploration of the microscopic mechanisms governing thermal transport in crystalline semiconductors is vital for optimizing and controlling high-performance thermoelectric devices.
\par To achieve low thermal conductivity in crystalline solids, various strategies have been proposed, including nanostructuring~\cite{plata2023harnessing}, atomic rattling~\cite{zheng2022effects}, strong lattice anharmonicity~\cite{chen2018high}, and liquid-like ions~\cite{liu2012copper}. For instance, the presence of rattlers results in a tenfold reduction in the phonon lifetime by enhancing the scattering phase space in clathrate  Ba$_8$Ga$_{16}$Ge$_{30}$(BGG)~\cite{tadano2018quartic}. While the strong anharmonic lattice vibration induced by the host-guest framework leads to significant anharmonic phonon scattering, this cage also facilitates electrical transport through a smooth channel and constitutes a PGEC system. Although low thermal conductivity is essential for high ZT, the importance of the intrinsic electronic structure cannot be underestimated or completely ignored in the pursuit of finding efficient thermoelectrics. It is becoming increasingly clear that complex band structures, including multi-valley Fermi surfaces~\cite{pei2011convergence}, convergence of bands~\cite{wang2014tuning, liu2012convergence}, and even coexistence of flat and dispersion bands~\cite{yue2023strong}, are key features of many good thermoelectric materials.
\par Copper-based chalcogenides have recently demonstrated advantages over traditional thermoelectric materials in industrial applications, primarily due to their nontoxicity, low cost, and environmental benignity~\cite{ma2020alpha}. Previously inspired by the synthesis of layered \textit{o}-CsCu$_5$S$_3$ for photovoltaic applications~\cite{xia2018cscu5se3, yang2021colloidal}, it has significantly advanced its role in the realm of thermoelectricity because of its high phase transition temperature of 823 K to \textit{t}-CsCu$_5$S$_3$. Notably, the \textit{o}-CsCu$_5$S$_3$ exhibits nearly the same intrinsically ultra-low $\kappa_\mathrm{L}$ of $<$ 0.6 Wm$^{-1}$K$^{-1}$ as the \textit{t}-CsCu$_5$S$_3$, while \textit{t}-CsCu$_5$S$_3$ has an extremely complex atomic configuration consisting of 72 atoms.~\cite{ma2019cscu5s3} More importantly, \textit{o}-CsCu$_5$S$_3$ displays a nonstandard glass-like temperature dependence—a rare characteristic in crystalline materials without impurities, disorder, and complex configurations.~\cite{simoncelli2019unified} Another demonstration of the potential of \textit{o-}CsCu$_5$S$_3$ for thermoelectric applications is the complete localization of all Cu atoms on their designated Wyckoff sites with full occupancy, which is fundamentally different from superionic phase compounds such as Cu-Q (Q=S, Se), AgCl, and AgBr.~\cite{zhang2020cu, xia2020high} For the superionic phase, the collisions between regular anion frames and dynamically changing cations strongly scatter phonons and suppress the $\kappa_\mathrm{L}$ to the glass-like limit.~\cite{wang2023theoretical, li2018liquid} 
In contrast, the strategic localization in \textit{o}-CsCu$_5$S$_3$ emerges as an effective preventive measure against device degradation.
Meanwhile, constructing a predictive theoretical model for the $\kappa_\mathrm{L}$ poses a significant challenge due to the necessity for an ab initio precision in characterizing interatomic interactions.~\cite{tong2023predicting} For one thing, this precision is required not only at the ground state (0 K) but also at finite temperatures, as the temperature dependence of phonon frequencies and anharmonic force constants is critical for accurately determining $\kappa_\mathrm{L}$ with strong anharmonic behavior.~\cite{xie2023microscopic} Moreover, the conventional particle-like approach to phonons may fall short in capturing the intricacies of thermal transport within the complex structure of \textit{o}-CsCu$_5$S$_3$. 
On the other hand, the implementation of four-phonon process computation presents a great challenge because of the larger cell size of o-CsCu$_5$S$_3$. This challenge stems from complex interactions during phonon scattering, which is crucial for accurately predicting the glass-like $\kappa_\mathrm{L}$.~\cite{han2022fourphonon} In particular, the possibility of a four-phonon process increases as the temperature increases, making the understanding of four-phonon scattering even more critical. Presently, \textit{o}-CsCu$_5$S$_3$ is not only devoid of comprehensive anharmonic phonon simulations but also lacks in-depth analysis of its electron configurations. This gap hinders our comprehension of the microscopic mechanisms underlying the anomalous $\kappa_\mathrm{L}$ and related thermoelectric characteristics.
\par In this work, we comprehensively investigate the thermal and electrical transport properties in crystalline \textit{o}-CsCu$_5$S$_3$ at the atomic level using first-principles density functional theory. Specifically, we integrate the state-of-the-art linearized Wigner transport equation (LWTE) and self-consistent phonon (SCP) theory to elucidate the thermal transport mechanisms in \textit{o}-CsCu$_5$S$_3$. By accounting for contributions to phonon energies and scattering rates from both cubic and quartic anharmonicities, we can well reproduce the experimental thermal conductivity and its temperature dependence. Meanwhile, the electrical transport properties are characterized by employing an accurate momentum relaxation-time approximation (MRTA) within the framework of the linearized Boltzmann transport equation (LBTE). Upon optimally tuning the carrier concentration, the theoretical maximum thermoelectric performance can achieve 18.4\% at 800 K, indicating its potential to meet critical criteria for thermoelectric materials. Our work elucidates the microscopic mechanisms behind the electronic and thermal transport characteristics in crystalline \textit{o}-CsCu$_5$S$_3$ and evaluates its practical application as a promising thermoelectric device.
\clearpage
\section{METHODS}
\par In this study, all density functional theory (DFT) calculations were carried out using the Vienna ab initio Simulation Package (VASP)~\cite{kresse1996efficient, kresse1996efficiency}. The project augmented wave (PAW) pseudopotentials were used to treat the Cs(5s$^2$5p$^6$6s$^1$), Cu(3d$^{10}$4s$^1$) and S(3s$^2$3p$^4$) shells as valence states. The Perdew-Burke-Ernzerhof functional (PBE) within the generalized gradient approximation (GGA) was employed for the exchange-correlation functional, and the optB86b-vdW was used to accurately describe the van der Waals (vdW) interactions~\cite{klimevs2011van}. A kinetic energy cutoff value of 600 eV and a 20×20×12 Monkhorst-Pack electronic \textit{k}-point grid were adopted to sample the Brillouin zone for the primitive cell. Tight energy and force convergence criteria of 10$^{-8}$ eV and 10$^{-6}$ eV·Å$^{-1}$, respectively, were employed for both structural optimization and static self-consistent DFT calculations.
\par  The zero-K harmonic (2$^{nd}$-order) force constants (IFCs) were extracted using the finite displacement approach~\cite{esfarjani2008method}. Static DFT calculations were performed using a 3×2×2 supercell containing 216 atoms, and a 4×4×4 \textit{k}-point meshes. To efficiently extract higher-order anharmonic IFCs, we constructed a displacement-force dataset with 200 atomic structures. Uniform displacements of 0.15 Å were enforced on all atoms along random directions using the random-seed method for reducing the cross-correlations between the atomic structures. The compressive sensing lattice dynamics (CSLD) approach ~\cite{zhou2019compressive} and the least absolute shrinkage and selection operator (LASSO) technique~\cite{nelson2013compressive}, implemented in the ALAMODE package~\cite{tadano2015self}, were employed to screen the physically important anharmonic terms of IFCs. After obtaining the harmonic and anharmonic IFCs, the anharmonic phonon energy renormalization was carried out by applying the self-consistent phonon theory considering both cubic and quartic anharmonicity in the reciprocal space~\cite{tadano2018quartic}. Within the linearized Wigner transport equation (LWTE) framework~\cite{simoncelli2019unified, simoncelli2022wigner}, the thermal conductivity $\kappa^\mathrm{P/C}_\mathrm{L}$ can be formulated as
\begin{equation}
    \begin{aligned}
\kappa^\mathrm{P/C}_\mathrm{L}=&\frac{\hbar ^2}{k_BT^2VN_0} \sum_{q}\sum_{j,j'}\frac{\Omega _{qj}+\Omega _{qj'}}{2} \upsilon _{qjj'} \otimes \upsilon _{qjj'} \\ 
&\times \frac{\Omega _{qj}n_{qj}(n_{qj}+1)+\Omega _{qj'}n_{qj'}(n_{qj'}+1)}{4(\Omega _{qj}-\Omega _{qj'})^2+ (\Gamma_{qj}+\Gamma_{qj'})^2}(\Gamma_{qj}+\Gamma_{qj'}) 
    \end{aligned}
\end{equation}
where the \textit{V} and \textit{N}$_0$ denote the primitive-cell volume and number of sampled, respectively. The $\Gamma_{q}$ stands for the scattering rates including three-phonon (3ph), four-phonon (4ph), and isotope-phonon scattering processes~\cite{xia2020microscopic, xia2020particlelike}. The particle-like phonon thermal transport properties are computed by using the ShengBTE~\cite{li2014shengbte} and FourPhonon packages~\cite{han2022fourphonon},
while the wave-like tunneling of phonon channels is evaluated using our in-house code.~\cite{zheng2024unravelling, zheng2022anharmonicity} (See supplementary materials for more details.)
\par The related electronic structure and transport calculations were carried out using the adjusted band gap obtained from HSE06, with spin-orbit coupling (SOC) included, employing the VASP and AMSET~\cite{ganose2021efficient} codes, respectively. The transition rates of electrons from the initial $\psi_{\textit{nk}}$ to final states $\psi_{\textit{mk+q}}$ based on Fermi's golden rule, which can be expressed as 
\begin{equation}    
\tilde{\tau}^{-1}_{\textit{nk} \rightarrow \textit{mk+q}} = \frac{2\pi}{\hbar}|g_{nm}(k,q)|^2\delta(\varepsilon_{nk}-\varepsilon_{mk+q})
\end{equation}
where $\varepsilon_{nk}$ symbolizes the specific energy state $\psi_{\textit{nk}}$. The $g_{nm}(k,q)$ accounts for three kinds of electron-phonon scattering matrix elements including acoustic deformation potential (ADP), polar optical phonon (POP), and ionized impurity (IMP) matrix element. The relaxation time of each electron can be evaluated by Matthiessen's rule, which is followed by
\begin{equation}    
\tau ^{-1}_{\textit{nk}}=\tau ^{-1}_{ADP}+\tau ^{-1}_{POP}+\tau ^{-1}_{IMP}
\end{equation}
\par Moreover, the net atomic charges and the overlap populations were obtained by using the density-derived electrostatic and chemical (DDEC6) method in Chargemol~\cite{manz2016introducing, limas2016introducing, manz2017introducing}. Meanwhile, the Multiwfn 3.8 program was employed to analyze the interactions by computing the interaction region indicators (IRI)~\cite{lu2021interaction}, which can be described as $\mathrm{IRI}(\textbf{r})=\left |  \bigtriangledown \rho (\textbf{r})\right | /[\rho (\textbf{r})]^a$ (Here the \textit{a} is a tunable parameter, we adopt a=1.1 for the standard definition of IRI). The correlative electronic wavefunction was achieved by Gaussian program~\cite{weedbrook2012gaussian}, and visualization was realized in the VMD 1.9.3 program, where the BGR color scale was adopted.
\clearpage
\section{RESULTS AND DISCUSSION}
\subsection{Crystal Configuration and Charge Distribution} 
\par The orthorhombic crystal structure of \textit{o}-CsCu$_5$S$_3$ is described by a space group of $Pmma$ (No. 55) with computed lattice parameters a = 3.896 Å, b = 8.857 Å, and c = 9.579 Å, as depicted in Figure 1a. Notably, these values are in good agreement with both experimental observations (a = 3.954 Å, b =  8.949 Å, and c = 9.636 Å) and other theoretical calculations~\cite{xia2018cscu5se3, chen2020beta}. The \textit{o}-CsCu$_5$S$_3$ structures are derived from the Cu$_4$S$_4$ columnar structural motif, which consists of the two Cu atoms with three-fold coordination labeled Cu(3) [see Figure 1b]. Specifically, these Cu$_4$S$_4$ units propagate with periodicity along [001] directions and form a wavy layer by extending from two opposite sides through the Cu atom with a two-fold coordination labeled Cu(2); meanwhile, an array of Cs$^+$ cations is accommodated between them, as shown in Figure 1a. The unique structural arrangement of \textit{o}-CsCu$_5$S$_3$ is expected to lead to a hierarchy of chemical bonds established within the material. The charge density equi-potential surface depicted in Figure S1 reveals that the overlapping charge cloud predominantly resides between the Cu and S atoms, contrasting with the non-overlapping charge spheres surrounding the Cs$^{+}$ions.
\par The comprehensive investigation of chemical bonding facilitates a deeper understanding of the physical properties exhibited by crystals. A sophisticated method for revealing interatomic interactions is the non-covalent interaction (NCL) index, which relies on the electron density $\rho$ and its derivatives~\cite{johnson2010revealing}. As depicted in Figure 1c, the NCI can be intuitively quantified by the interatomic region indicator (IRI) based on optimized reduced density gradients (RDG), a function of sign($\lambda_2$)$\rho$, where sign($\lambda_2$) indicates the sign of the second eigenvalue of the electron density Hessian matrix~\cite{contreras2011nciplot}. Generally, negative value of $\mathrm{sign} (\lambda_2$)$\rho$ indicates attractive (or bonding) interactions. In regions of low electron density, weak interactions typically manifest as a spike, characterized by a significant change in RDG near the zero point.~\cite{lu2021interaction} In Figure 1c, van der Waals (vdW) interactions, approaching zero ( $\left | \mathrm{sign} (\lambda_2)\rho \right |  < $ 0.01 a.u.), are observed at the critical point within the low electron density region. Meanwhile, there are also numerous bonding interactions exhibiting tendencies towards weak covalent character( $\mathrm{sign} (\lambda_2)\rho < $ -0.02 a.u.). In addition, Figure 1d illustrates 3D IRI isosurfaces with BGR color scales representing the sign($\lambda_2$)$\rho$ values.  As expected, the 2D [Cu$_5$S$_3$]$^{-}$ and Cs$^{+}$ layers are predominantly interspersed with weak van der Waals interactions, which facilitate the stochastic oscillation of Cs atoms. In addition, the intralayer Cu–S interactions exhibit a strong covalent character within the [Cu$_5$S$_3$]$^{-}$ slab, while the red area surrounding the bonding reveals the existence of steric effects. 
\subsection{Electronic States and Anharmonically Renormalized Phonons}
\par \noindent \textbf{Electronic band structure, DOS, and bonding states.} 
In Figure 2a-c, we illustrate the atom-projected electronic band structures, the projected density of states (PDOS), and the crystal orbital Hamiltonian populations (COHP) for \textit{o}-CsCu$_5$S$_3$. Our calculated indirect band gap of 1.28 eV places the valence band maximum (VBM) at the Y point and conduction band minimum (CBM) at the $\Gamma$ points, consistent with the experimental observation [see Figure 2a].~\cite{chen2020beta} We next move on to conduct an in-depth analysis of PDOS and COHP to delve into the stereochemical interactions among atomic pairs. Near the Fermi level, the valence band is primarily characterized by Cu 3$d$ orbitals and S 4$p$ orbitals, as shown in Figure 2b. This notable $p$-$d$ orbital hybridization can be ascribed to the relatively high-energy $d$-states in copper, which closely match the energy levels of the S $p$-states.~\cite{jaffe1984theory} 
In particular, we note that the main contribution of Cu atoms near the Fermi surface comes from Cu(3) atoms, which means that the electron transport properties are dominated by Cu(3) compared to Cu(2). [see Figure S2] The Cs atoms, on the other hand, are primarily bonded to the [Cu$_5$S$_3$]$^{-}$ layers through weak van der Waals interactions [See Figures 1a and 1c-d]. As a result, the electrons associated with Cs atoms are tightly bonded to their respective nuclei, leading to deeper energy levels that do not contribute to the electronic states at the Fermi surface. Hence, Cs atoms make a negligible contribution to the electronic properties at the Fermi level [see Figure 2a-b]. Indeed, the COHP plot depicted in Figure 2c reveals that the Cs atoms exhibit minimal orbital overlap with Cu and S atoms due to their spatial localization, leading to an absence of any bonding or antibonding states between them. On the contrary, the $d$-orbitals of Cu atoms are energetically close to the $p$-orbitals of S, which facilitates a robust $p$-$d$ hybridization, as shown in Figure 2c. This robust hybridization elevates the filled antibonding states near the Fermi level, thereby weakening the bond strength and forming metavalent-like bonding characteristics.~\cite{gholami2024unlocking} In particular, the occupation of antibonding states in the upper valence band serves as the origin of strong lattice anharmonicity and structural instability.
\par \noindent \textbf{Anharmonic lattice dynamics.} In functional materials, both the cubic and quartic anharmonicities are essential for the accurate modeling of lattice dynamics.~\cite{zheng2022effects, zheng2022anharmonicity} For crystalline \textit{o}-CsCu$_5$S$_3$, 0-K phonon dispersions and the anharmonically renormalized phonon dispersions arising from both cubic and quartic anharmonicity at finite temperatures are illustrated in Figure 2d-e. Obviously, no imaginary frequencies are observed in zero-K phonon dispersions, indicating the dynamical stability of \textit{o}-CsCu$_5$S$_3$. When considering the effect of three- (3ph) and four-phonon (4ph) interactions on phonon dispersions, distinctive phonon energy shifts were observed in \textit{o}-CsCu$_5$S$_3$. Although the low-frequency optical modes undergo slight hardening or softening, the high-frequency optical modes experience significant softening. It is worth noting that the softening phenomenon in high-frequency optical modes has also been observed in other highly anharmonic compounds, including the CsCu$_2$I$_3$~\cite{zheng2023wave} and BaZrO$_3$~\cite{zheng2022anharmonicity}.
However, considering only the effect of quartic anharmonicity, i.e., loop self-energy, on phonon energy shifts may lead to an overestimation issue in highly anharmonic materials~\cite{tadano2022first}. In Figure S3, a notable hardening trend is observed across the almost entire phonon spectrum with increasing temperature. However, accounting for the effect of cubic anharmonicity, i.e., bubble self-energy, on phonon energy shifts dramatically suppresses the phonon hardening, owing to negative frequency shifts from the bubble diagram.  Therefore, the competition between the quartic anharmonicity and the cubic anharmonicity accounts for the predicted phonon softening of \textit{o}-CsCu$_5$S$_3$. This observation highlights the crucial role of cubic anharmonicity (bubble diagram) in accurately predicting finite-temperature phonon energies in \textit{o}-CsCu$_5$S$_3$.~\cite{zheng2022anharmonicity} More importantly, an accurate lattice dynamics modeling is essential for investigating the properties of thermal transport, which will be discussed later. On the other hand, the dispersive low-lying optical (LLO) branch manifests avoided crossings with the longitudinal acoustic (LA) branch, as the indication of phonon polarization hybridization, as shown in Figure S4.~\cite{christensen2008avoided} Traditionally, the phonon group velocities near the avoided crossing points are suppressed, thereby constraining the transport efficiency of heat-carrying phonons~\cite{li2022high}. Other research has pointed out that optical-acoustic phonon coupling can also enhance scattering rates and suppress lattice thermal conductivity through the hybridization of optical phonon eigenvectors into acoustic phonons~\cite{li2016influence}. Meanwhile, numerous flattened low-frequency optical branches with soft vibrations act as Einstein oscillators, which have also been demonstrated to be advantageous for phonon scattering~\cite{li2023wavelike, liu2016reduction}. These indices anticipate the existence of strong lattice anharmonicity, which can severely suppress the heat transport in the \textit{o}-CsCu$_5$S$_3$ crystals. 

\par In Figures 3a-d, we present color-coded visualizations of the participation ratios associated with different atoms, providing a direct representation of their contributions to the phonon dispersions. Obviously, the low-frequency phonons with frequencies below 60 cm$^{-1}$ in \textit{o}-CsCu$_5$S$_3$ are predominantly dominated by the Cs atoms due to their heavy mass and weak bonding [see Figures 2d-e and 3a]. Conversely, the S atoms mainly contribute to the high-frequency optical branches due to their relatively small atomic mass. Nevertheless, the influence of copper atoms on phonon dispersions is more complex, given their involvement with two coordinated copper atoms. Specifically, the Cu(3) atoms significantly contribute to the entire low to mid-frequency phonon spectrum, highlighting its critical role in thermal transport in \textit{o}-CsCu$_5$S$_3$, where low-frequency phonons act as the main heat carriers [see Figure 3c]. In contrast, the Cu(2) atoms exhibit a participation ratio primarily localized within the mid-frequency domain, resulting in a less significant impact on thermal transport compared to that of the Cu(3) atoms [see Figure 3d]. We attribute the dominance of Cu(3) atoms on the low-frequency phonons to the presence of antibonding states, arising from the strong $p$-$d$ orbital hybridization between Cu and S atoms.~\cite{gholami2024unlocking} Furthermore, we present the mean squared atomic displacements (MSD) for all atoms, offering valuable insights into the dynamics of atomic vibrations in Figure S5. In addition to the large MSD of Cs atoms, quantified at 0.021 Å$^2$, the Cu atoms within both complexes also demonstrate substantial MSDs, with values of 0.019 Å$^2$ and 0.018 Å$^2$ respectively. These values are larger than those observed in other Cu-based crystals with strong anharmonicity, such as CuSbS$_2$ (0.015 Å$^2$)~\cite{feng2017dual}, further highlighting the role of anti-bonding states in inducing the softening of Cu-S bonds.

\subsection{Dual-channel thermal transport model}
\par \noindent \textbf{Lattice thermal conductivity.} We proceed by calculating the total $\kappa_\mathrm{L}$, which incorporates population conductivity $\kappa^\mathrm{P}_\mathrm{L}$ from particlelike propagation channel and coherence conductivity $\kappa^\mathrm{C}_\mathrm{L}$ from wavelike tunneling channel, within the framework of the Wigner formalism of quantum mechanics~\cite{simoncelli2019unified, simoncelli2022wigner}. The $\kappa^\mathrm{P}_\mathrm{L}$ corresponds to the conventional Peierls-Boltzmann thermal conductivity in crystals, which is derived from the diagonal terms ($j = j'$) of the heat-flux operator. When phonon branches are well-separated, thermal transport is dominated by a particle-like phonon propagation channel, as demonstrated by Tl$_3$VSe$_4$.~\cite{xia2020particlelike}  When the phonon mean free path (MFP) approaches the interatomic spacing, the off-diagonal terms of heat flux operators ($j \neq j'$)  become significant and give rise to $\kappa^\mathrm{C}_\mathrm{L}$, as observed in double perovskite Cs$_2$AgBiBr$_6$.~\cite{zheng2024unravelling} Here, using the advanced thermal transport model, i.e., the SCPB+3,4ph+OD, considering phonon scatterings and energy shifts from both 3ph and 4ph processes and diagonal/non-diagonal terms of heat flux operators, allows us to obtain reliable $\kappa_\mathrm{L}$. In Figure 4a, the converged $\kappa_\mathrm{L}$ at 300 K along the $x$- and $z$-directions reaches exceptionally low values of 0.46 and 0.38 Wm$^{-1}$K$^{-1}$, respectively, in good agreement with the experimental values (0.35 $\sim$ 0.50 Wm$^{-1}$K$^{-1}$).~\cite{ma2019cscu5s3} Along its stacking direction, the calculated $\kappa_\mathrm{L,y}$ reaches an exceptionally low value of 0.150 Wm$^{-1}$K$^{-1}$, as depicted in Figure S6, merely five times that of air (0.025 Wm$^{-1}$K$^{-1}$ at 300 K). This can be attributed to the small group velocity induced by the weak vdW interactions between the sublattice [Cu$_5$S$_3$]$^{-}$ and Cs$^{+}$ atoms [see Figure 1c-d]. Meanwhile, our prediction also well reproduces the experimentally measured temperature dependence of $\kappa_\mathrm{L}$ along with the parallel to the Bridgman growth direction ($T^{-0.33}$) [see Figure 4b]. 
\par However, when considering only 3ph scattering processes in \textit{o}-CsCu$_5$S$_3$ (SCP+3ph model), the predicted $\kappa_\mathrm{L}$ is $\sim$ 0.62 Wm$^{-1}$K$^{-1}$ at room temperature, obtained by averaging values along the two principal crystallographic axes. Meanwhile, the predicted $\kappa_\mathrm{L}$ follows a temperature dependence of $T^{-0.55}$ in the temperature range of 300-800 K [see Figure 4b]. Further including 4ph scatterings, i.e., SCP+3,4ph model, results in significant discrepancies between the experiment and prediction in thermal conductivity, especially at high temperatures [see Figure 4b]. Considering the positive temperature dependence of $\kappa^\mathrm{C}_\mathrm{L}$, adding the coherence contribution to the predicted $\kappa_\mathrm{L}$ from above model will result in an overestimation in thermal conductivity at low temperatures.~\cite{simoncelli2019unified, simoncelli2022wigner} Recent studies have highlighted the significance of negative phonon energy shifts in accurately reproducing phonon energy and thermal conductivity at finite temperatures.~\cite{tadano2022first, tong2020first} Therefore, we integrate the negative phonon energy into the SCP+3ph and SCP+3,4ph thermal transport models mentioned above. In Figure 4b, both models, i.e., SCPB+3ph and SCPB+3,4ph models, underestimate the lattice thermal conductivity compared with experimental data throughout the entire temperature range. As discussed previously, the SCPB+3,4ph+OD model, incorporating the additional contribution from the wave-like phonon tunneling channel, accurately reproduces the experimental thermal conductivity. Specifically, using SCPB+3,4ph model, the predicted $\kappa^\mathrm{P}_\mathrm{L}$ values are 0.267 and 0.067 W/mK at 300 and 800 K, respectively. Correspondingly, the predicted $\kappa^\mathrm{C}_\mathrm{L}$ values are 0.157 and 0.222 W/mK at 300 and 800 K, respectively, accounting for 37\% and 76\% of the total $\kappa^\mathrm{P}_\mathrm{L}$. Therefore, the critical or dominant role of coherence conductivity gives rise to the nonstandard extremely weak temperature dependence of total $\kappa_\mathrm{L}$ in \textit{o}-CsCu$_5$S$_3$. Consequently, the state-of-the-art heat transport theory used in this work is essential for reproducing the experimental temperature dependence of $\kappa_\mathrm{L}$, especially in glass-like and disordered crystals~\cite{shenogin2009predicting}.

\par \textbf{Particle-like phonon transport properties.} To unveil the microscopic mechanisms of thermal transport in \textit{o}-CsCu$_5$S$_3$, we present the phonon lifetime $\tau(\mathbf{q})$ as a function of phonon frequency at 300 K and 800 K, respectively in Figure 4c-d. Notably, nearly all phonons exceed the  Ioffe-Regel limit, i.e., $\tau(q) = [\Gamma (q)]^{-1} > 1/\omega (q)$, validating the theoretical framework for describing transport using the phonon quasiparticle excitation concept.~\cite{simoncelli2022wigner} Additionally, we observe that the phonons with frequencies below 150 cm$^{-1}$ contribute to large phonon scattering rates, which is consistent with the phonon frequency region dominated by Cu(3) and Cs atoms. Furthermore, the scattering rates due to four-phonon processes are comparable to or surpass the scattering rates due to three-phonon processes at both 300 K and 800 K. [see Figure S7] In particular, the peak region of the 4ph scattering process in \textit{o-}CsCu$_5$S$_3$ coincides with the partial APR of Cs atoms [see Figure 3a], which indicates that Cs atoms play an important role in the strong 4ph scattering rates. This observation emphasizes the necessity of 4ph scatterings in accurately evaluating thermal transport in highly anharmonic \textit{o}-CsCu$_5$S$_3$. The timescale $[\Delta \omega_{av}]^{-1}$, known as the "Wigner time limit," can distinguish the relative contributions of particle-like and wave-like thermal transport mechanisms~\cite{simoncelli2022wigner}. Phonons with $\tau_\mathrm{Ioffe} < \tau < \tau_\mathrm{Wigner}$ mainly influence $\kappa^\mathrm{C}_\mathrm{L}$, while those with $\tau > \tau_\mathrm{Wigner}$ primarily contribute to $\kappa^\mathrm{P}_\mathrm{L}$. From Figures 4c-d, we observe that many phonons have lifetimes below the Wigner time limit, indicating the necessity of invoking the wave-like tunneling transport channel in \textit{o}-CsCu$_5$S$_3$, irrespective of whether 4ph scatterings are considered.

\par To gain a deeper insight into the thermal transport in \textit{o}-CsCu$_5$S$_3$, we examine the population conductivity spectrum $\kappa^\mathrm{P}_\mathrm{L}(\omega)$ and corresponding cumulative, as shown in Figure 5. The ultra-low $\kappa^\mathrm{P}_\mathrm{L,\textit{x}}$ and $\kappa^\mathrm{P}_\mathrm{L,\textit{z}}$ are primarily dominated by acoustic and low-frequency optical modes ($<$ 130 cm$^{-1}$). Considering the low-frequency phonons dominated by Cu(3) atoms [see Figure 3c], we attribute the ultra-low $\kappa^\mathrm{P}_\mathrm{L}$ to the presence of antibonding states resulting from S(p)-Cu(d) orbital hybridization. Similar phenomena have also been observed in double perovskite Cs$_2$AgBiBr$_6$~\cite{zheng2024unravelling}, where the Br-dominated low-frequency phonons act as depressors for population conductivity. Furthermore, as depicted in Figure 5a-d, we observe that the significant reduction in $\kappa^\mathrm{P}_\mathrm{L}$ due to 4ph scatterings is primarily contributed by phonons with frequencies below $\sim$ 50 cm$^{-1}$. This reduction in  $\kappa^\mathrm{P}_\mathrm{L}$ can be mainly attributed to the rattling-like flat phonon modes dominated by Cs atoms due to weak bonding, similar phenomena have also been observed in materials, such as the crystalline AgCrSe$_2$ and Cs$_3$Bi$_2$I$_6$Cl$_3$~\cite{li2023wavelike, xie2020first}. Overall, the presence of antibonding states from Cu(3)-S pairs, coupled with the stochastic oscillation of Cs atoms, drives the population conductivity towards the amorphous limit in \textit{o}-CsCu$_5$S$_3$.

\par \textbf{Wave-like phonon transport properties.} To better understand the microscopic mechanisms of wave-like tunneling transport, we calculate the spectral and cumulative $\kappa^\mathrm{C}_\mathrm{L,\textit{x}}$ and $\kappa^\mathrm{C}_\mathrm{L,\textit{z}}$ for \textit{o}-CsCu$_5$S$_3$ at 300 and 800 K, respectively. In Figures 6a-b, coherence conductivity is predominantly driven by phonons with frequencies below 150 cm$^{-1}$, a phenomenon mainly attributed to the strong anharmonicity originating from Cu(3) and rattling-like Cs atoms and dense inter-branch spacing [see Figure 3b-c]. With increasing temperature, the $\kappa^\mathrm{C}_\mathrm{L,\textit{x,z}}$ also increases due to the excitation of high-frequency phonons.~\cite{tadano2018quartic} Therefore, the strong anharmonicity results in the dominant role of coherence conductivity in \textit{o}-CsCu$_5$S$_3$ at high temperatures. To visualize the coherence contributions more intuitively, we present the two-dimensional $\kappa^\mathrm{C}_\mathrm{L,\omega{\textit{jj'}}}$ for the coherence conductivity at 300 K and 800 K, as demonstrated in Figure 6c-f. Notably, the quasi-degenerate vibrational eigenstates $(\omega_{qj} \cong  \omega_{qj})$ predominantly contribute to coherence conductivity along both the \textit{x}- and \textit{z}-directions at 300 K [see Figures 6c-d].~\cite{simoncelli2019unified} At 800 K, non-degenerate phonons begin to contribute to coherence conductivity due to the thermal activation of high-frequency phonons, especially along the \textit{x}-direction, where they can even become dominant [see Figures 6e-f]. This observation emphasizes the importance of considering coupled vibrational eigenstates to accurately elucidate the origin of anomalous lattice thermal transport.

\subsection{Electronic transport and thermoelectric performance}
\par \noindent \textbf{Electronic transport and properties.} The experimentally and theoretically observed exceptionally low thermal conductivity $\kappa_\mathrm{L}$ in \textit{o}-CsCu$_5$S$_3$ highlights its promising potential for thermoelectric application. The electronic transport properties of \textit{o}-CsCu$_5$S$_3$ can be rigorously evaluated by conducting a theoretical analysis of its electronic band structure. As previously mentioned, the presence of antibonding states below the Fermi level weakens the Cu-S bonds, influencing both the valence band characteristics and the electron transport properties. Within this framework, two distinct band characteristics emerge the 'pudding-mold,' which is characterized by the coexistence of flat and dispersive bands, and the presence of high-degeneracy valleys in \textit{o}-CsCu$_5$S$_3$.~\cite{gholami2024unlocking} The electronic bands in \textit{o}-CsCu$_5$S$_3$ exhibit flat and dispersive characteristics along the $\Gamma$–Z, Y–U, and Y-S directions, respectively [see Figure 2a]. The band structure characteristics in \textit{o}-CsCu$_5$S$_3$ meet the criteria of good thermoelectric materials proposed by Sofo and Mahan.~\cite{mahan1996best} The "pudding-mold" band characteristics contribute to concentrated energy distribution in the "flat" band, while maintaining high carrier mobility within the "dispersive" bands, resulting in a high power factor PF.~\cite{isaacs2018inverse} 
Additionally, the presence of multiple valleys with high degeneracy ($N_v$) in the first Brillouin zone, where $N_v$ represents the total number of carrier pockets, also benefits the thermoelectric properties.~\cite{tan2017improving} In \textit{o}-CsCu$_5$S$_3$, the energy difference between the highest point of the valence band at Y and the secondary valence band (VBM2) at $\Gamma$ is a small value of 0.18 eV. As mentioned above, the energy difference between PF and the valence band extreme $\Delta E$ is exponentially dependent through PF $\propto$ $e^{-\Delta E/k_BT}$, reaching its maximum at $\Delta E \approx 0$.~\cite{diznab2019achieving} Therefore, the high degenerate valence valleys significantly increase electron states near the Fermi surface, thereby improving the power factor and thermoelectric performance.~\cite{pei2011convergence, tan2017improving} 
\par The theoretically observed high-lying valence band maximum and low ionization potential in \textit{o}-CsCu$_5$S$_3$ play a crucial role in facilitating the formation of hole carriers, a phenomenon demonstrated by prior experiments~\cite{ma2019cscu5s3}. Therefore, fine-tuning the hole concentration can further enhance the attainment of optimal thermoelectric properties in \textit{o}-CsCu$_5$S$_3$. Figure 7 illustrates the temperature-dependent characteristics of \textit{p}-type electrical conductivity ($\sigma$), Seebeck coefficient ($S$), power factor (PF), and electronic thermal conductivity ($\kappa_e$) for \textit{o}-CsCu$_5$S$_3$. These electronic transport properties are computed across a range of charge carrier concentration from 10$^{18}$ to 10$^{21}$ cm$^{-3}$. Generally, the Cu-S bonds with the wider orbital overlap tend to facilitate the carrier transitions and ensure larger conductivity.~\cite{yue2024role} Among the carrier scattering mechanisms, we note that the scattering of polar optical phonons to carriers has the largest perturbation of charge carriers across the whole range, which severely limits the mobility of electrons [see Figure S8]. In addition, an environment with lower temperatures and higher carrier concentrations ensures that maximum conductivity ($\sim$21.9 $\times$ 10$^4$ Sm$^{-1}$) is achieved as the lower temperatures can reduce the scattering of carriers while higher concentrations enhance the overall electrical transport capacity. Meanwhile, the electronic thermal conductivity $\kappa_e$ is positively correlated with the conductivity $\sigma$, which can be described as Wiedemann-Franz law~\cite{lavasani2019wiedemann} ($\kappa_e = L\sigma T$, where \textit{L} is the Lorenz number). Accordingly, they show a similar distribution on the heat map. For Seebeck coefficient $S$, it usually shows the opposite trend to electrical conductivity, therefore, its maximum value reaches 672 $\mathrm{\mu VK^{-1}}$ at $10^{21}$ cm$^{-3}$ and 800 K. Finally, the theoretical maximum power factor is achieved with 1.48 mWm$^{-1}$K$^{-1}$ at 800 K and carrier concentration of 3.86$\times$10$^{20}$ cm$^{-3}$ under the competitive relationship between electron transport parameters. 
\par \noindent \textbf{Thermoelectric conversion efficiency.}
To achieve the maximum thermoelectric value ZT, it is imperative to intricately balance the interrelationships among various electronic transport parameters. In Figure 8a, the peak ZT values for \textit{o}-CsCu$_5$S$_3$ are observed at 800 K, reaching a maximum of 1.88, which corresponds to an optimal charge carrier concentration of 5.74$\times$10$^{19}$ cm$^{-3}$. Note that the intrinsic ZT value of o-CsCu$_5$S$_3$, without considering doping, was experimentally measured to be 0.46 at 800 K.~\cite{ma2019cscu5s3} However, our computational analysis is confined to evaluating the electronic transport properties at select doping concentrations. This assessment omits the specification of particular p-type dopants, thereby implying a broader scope for further experimental and theoretical investigations. Such explorations are imperative for a comprehensive understanding of dopant-dependent transport phenomena and for optimizing performance for practical applications.
Assuming that the required optimal carrier concentrations can be achieved through doping, \textit{p}-type \textit{o}-CsCu$_5$S$_3$ emerges as a highly promising material for high-temperature thermoelectric applications, given its high ZT values and phase transition temperature. In the realm of thermoelectric materials, the conversion efficiency $\eta$ is recognized as a practical indicator for evaluating thermoelectric performance. Typically, a conversion efficiency of no less than 15\% is required for thermoelectric devices to meet the crucial criterion for market applications~\cite{liu2017new}. Figure 8b illustrates the conversion efficiency $\eta$, wherein the hot end temperature spans from 400 to 800 K, and the cold end temperature defaults to 300 K. The optimal conversion efficiency $\eta$ of 18.4\% is achieved at the same temperature and carrier concentration corresponding to the maximum ZT value. This serves as additional confirmation of its potential practical application value.
\clearpage
\section{CONCLUSIONS}
\par In conclusion, we have systematically investigated the anharmonic lattice dynamics, microscopic mechanisms of electronic, and thermal transport, as well as the thermoelectric properties in crystalline \textit{o}-CsCu$_5$S$_3$. Our findings demonstrate that the loose Cu–S bonding, induced by the antibonding state through $p$-$d$ hybridization, combined with the stochastic oscillation of Cs atoms, leads to strong anharmonicity in \textit{o-}CsCu$_5$S$_3$. This observation emphasizes the critical importance of cubic and quartic anharmonicity in accurately modeling anharmonic lattice dynamics and evaluating thermal transport properties. The advanced thermal transport model, namely the SCPB+3,4ph+OD model, incorporating phonon energy shifts and scattering rates from both 3ph and 4ph processes, as well as diagonal and non-diagonal terms of heat flux operators, is employed to predict thermal transport in \textit{o-}CsCu$_5$S$_3$. Specifically, we can well reproduce the experimentally measured ultra-low thermal conductivity of 0.42 W/mK and its glassy temperature dependence, following $T^{-0.33}$. Additionally, the presence of low-frequency modes dominated by Cu(3) atoms, along with flat modes contributed by Cs atoms, drives the population conductivity to its amorphous limit in \textit{o-}CsCu$_5$S$_3$. This leads to a transformation in heat conductance from particle-like phonon propagation to wavelike phonon tunneling transport, ultimately dominating heat transfer at high temperatures. Moreover, the presence of strong antibonding states between the Cu(3)-S pairs near the Fermi energy level not only weakens the chemical bonding but also affects the characteristics of the valence band and the behavior of the electrons. In particular, the "pudding mold" band shape—exhibiting a flat-and-dispersive structure—alongside the highly degenerate valence band, facilitates the achievement of a high p-type power factor. Using an accurate momentum relaxation-time approximation (MRTA) approach, at an optimal hole concentration of 5.74$\times$10$^{19}$ cm$^{-3}$, \textit{p}-type \textit{o}-CsCu$_5$S$_3$ exhibits a maximum thermoelectric value ZT of 1.88 and a conversion efficiency of 18.4\% at 800 K. This observation confirms the potential advantages of crystalline in \textit{o}-CsCu$_5$S$_3$ the filed of thermoelectric materials. Our work not only accurately unravels the correlation between ultra-low glassy thermal transport, electronic transport, and crystal structure in \textit{o}-CsCu$_5$S$_3$, but also paves the way to better understand, design, and manipulate other thermoelectric materials.

\clearpage
\section*{Supporting Information}
Total charge-density distribution within the primitive cell; Projected electronic band structures of Cu(2) and Cu(3) under the same weight; Anharmonically renormalized phonon dispersions at finite temperatures; Phonon band dispersion in the low-frequency region; Mean square atomic displacements (MSDs) with temperature; Variations of temperature-dependent thermal conductivity along with y axis; Scattering rates (SRs) of three- and four-phonon processes; Scattering rates as a function of temperature at specific hole carrier concentration 

\section*{Author Contributions}
\par \noindent The authors confirm their contribution to the paper as follows: writing - original draft $:$ J.C.Y. and J.Z.Z.; conceptualization $:$ J.Z.Z; data curation $:$ J.C.Y., J.D.L., and W.L.R; investigation$:$ S.Q.G and H.L.; writing - review $\&$ editing $:$ Y.H.L and T.C.; funding acquisition $:$ T.C. All authors reviewed the results and approved the final version of the manuscript.
\section*{Conflicts of interest}
\par \noindent There are no conflicts to declare.

\begin{acknowledgement}
\par \noindent This work was supported by the National Natural Science Foundation of China (Grant No. 52072188, No. 12204254), the Program for Science and Technology Innovation Team in Zhejiang (Grant No. 2021R01004), 
and the Natural Science Foundation of Zhejiang province (Grant No. LQ23A040005). We are grateful to the Institute of High-pressure Physics of Ningbo University for its computational resources.
\end{acknowledgement}
%%%%%%%%%%%%%%%%%%%%%%%%%%%%%%%%%%%%%%%%%%%%%%%%%%%%%%
\bibliography{achemso-demo}

\clearpage
\begin{figure}[h]
\centering
  \includegraphics[height=12.5cm]{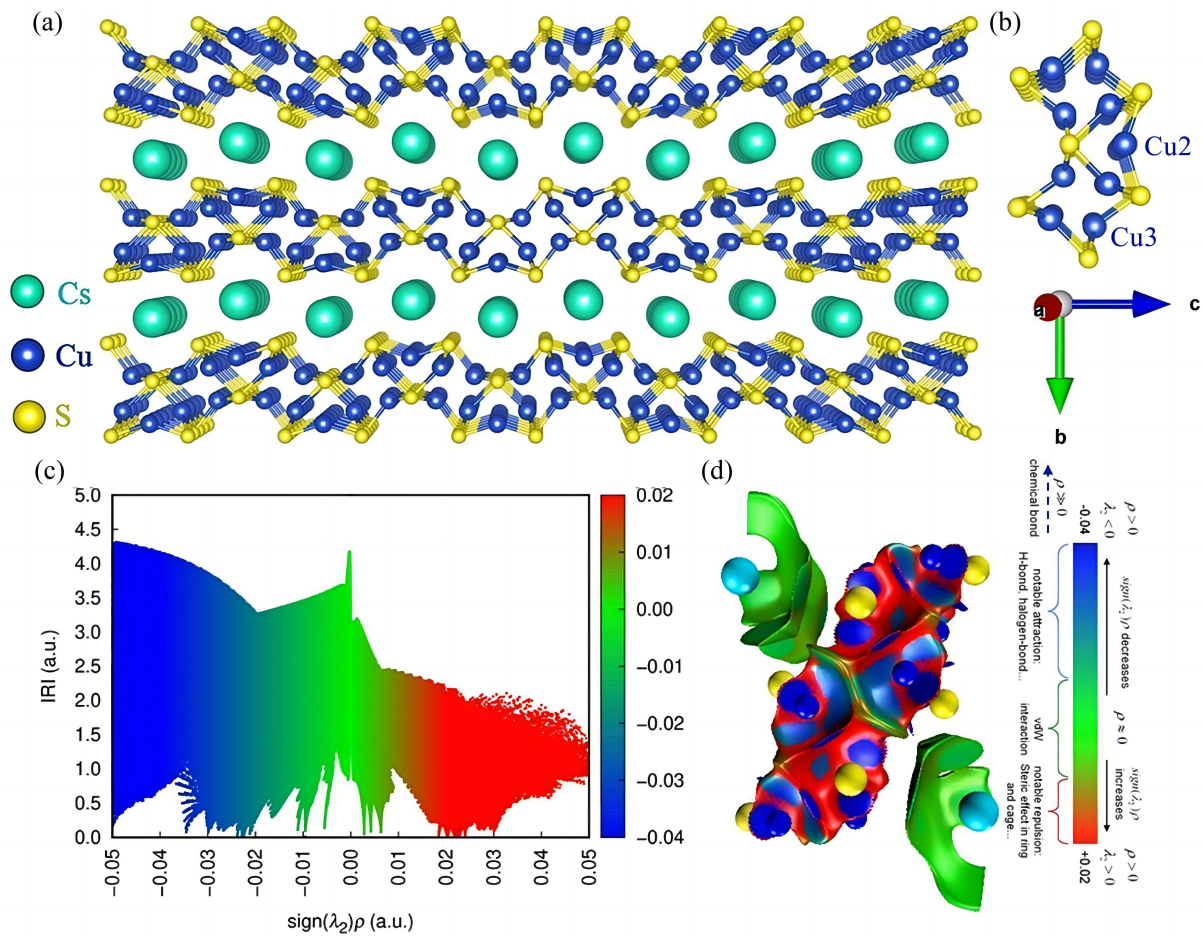}
  \caption{(a) The crystal structure of crystalline \textit{o}-CsCu$_5$S$_3$, featured by wave-shaped Cu$_4$S$_4$ column units with interspersed Cs atoms. The cyan, blue, and yellow spheres represent cesium, copper, and sulfur atoms, respectively. (b) The Cu$_4$S$_4$ columnar structure building unit constructed by 3-fold coordinated Cu(3) and 2-fold coordinated Cu(2) atoms. (c) Non-covalent interaction analysis based on interaction region indicator (IRI).
  Negative and positive sign($\lambda_2$)$\rho$ denotes attractive and repulsive interactions, respectively, and weak interactions are pervasive in the vicinity of sign($\lambda_2$)$\rho$ = 0. (d) Isosurface map with standard coloring method and chemical explanation of sign($\lambda_2$)$\rho$ on IRI isosurfaces. The blue, green, and red areas represent covalent, van der Waals, and repulsive interactions, respectively.}
  \label{Fig1}
\end{figure}
\clearpage
\begin{figure}[t]
\centering
%\begin{center}
    \includegraphics[height=16cm]{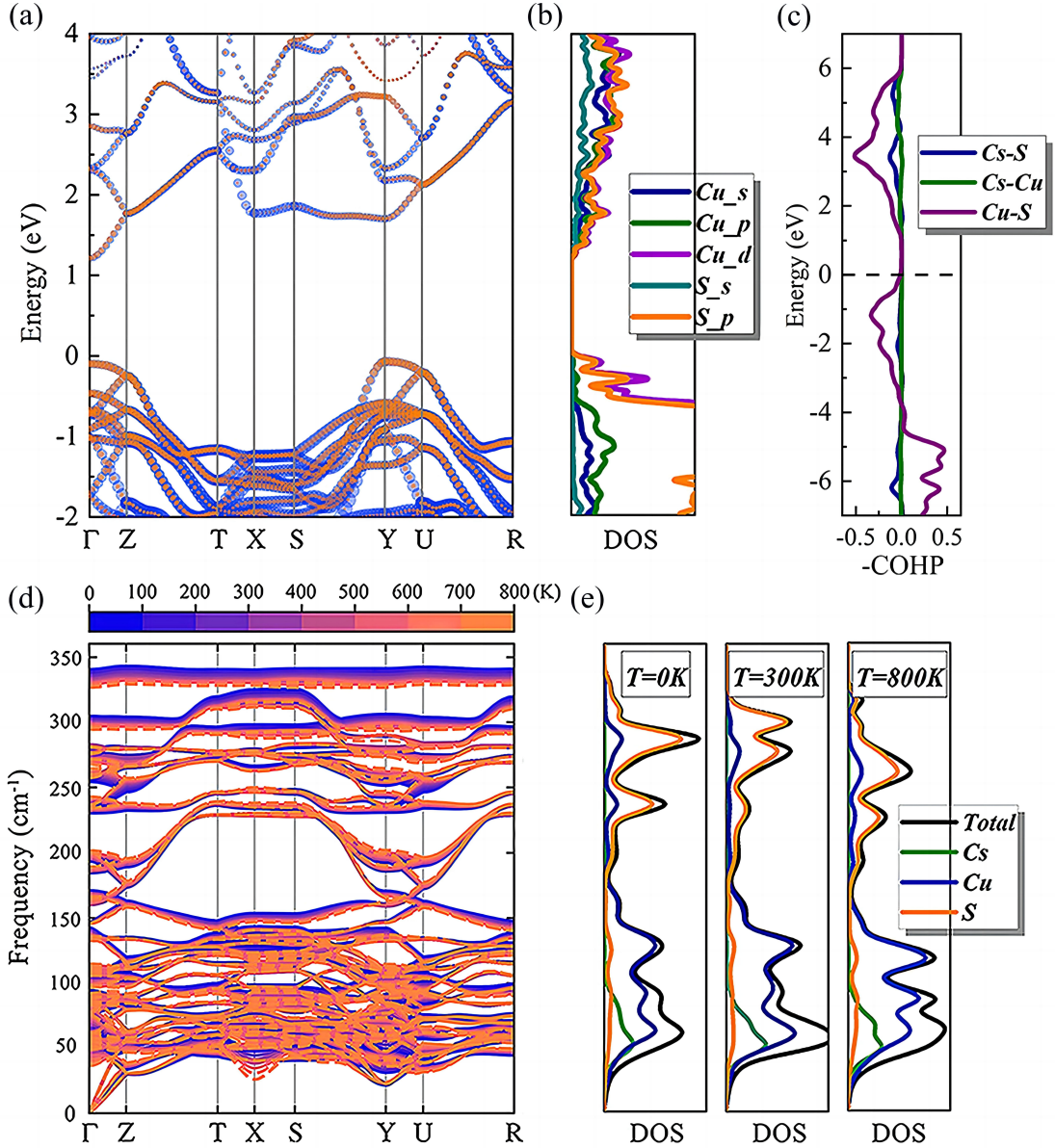}
  \caption{(a) Atom-projected electronic band structures, (b) the projected density of states, and (c) the COHP plots for \textit{o}-CsCu$_5$S$_3$. The blue and orange projections represent the contributions of Cu and S atoms, respectively. (d) Calculated anharmonically renormalized phonon dispersions at finite temperatures and compared with the zero-K phonon dispersions obtained by using harmonic approximation. (e) The right panel shows the decomposed-atom partial and total phonon densities of states (DOS) at 0 K, 300 K, and 800 K, respectively.}
  \label{Fig2}
%\end{center}
  
\end{figure}
\clearpage
\begin{figure}[t]
\centering
  \includegraphics[height=12cm]{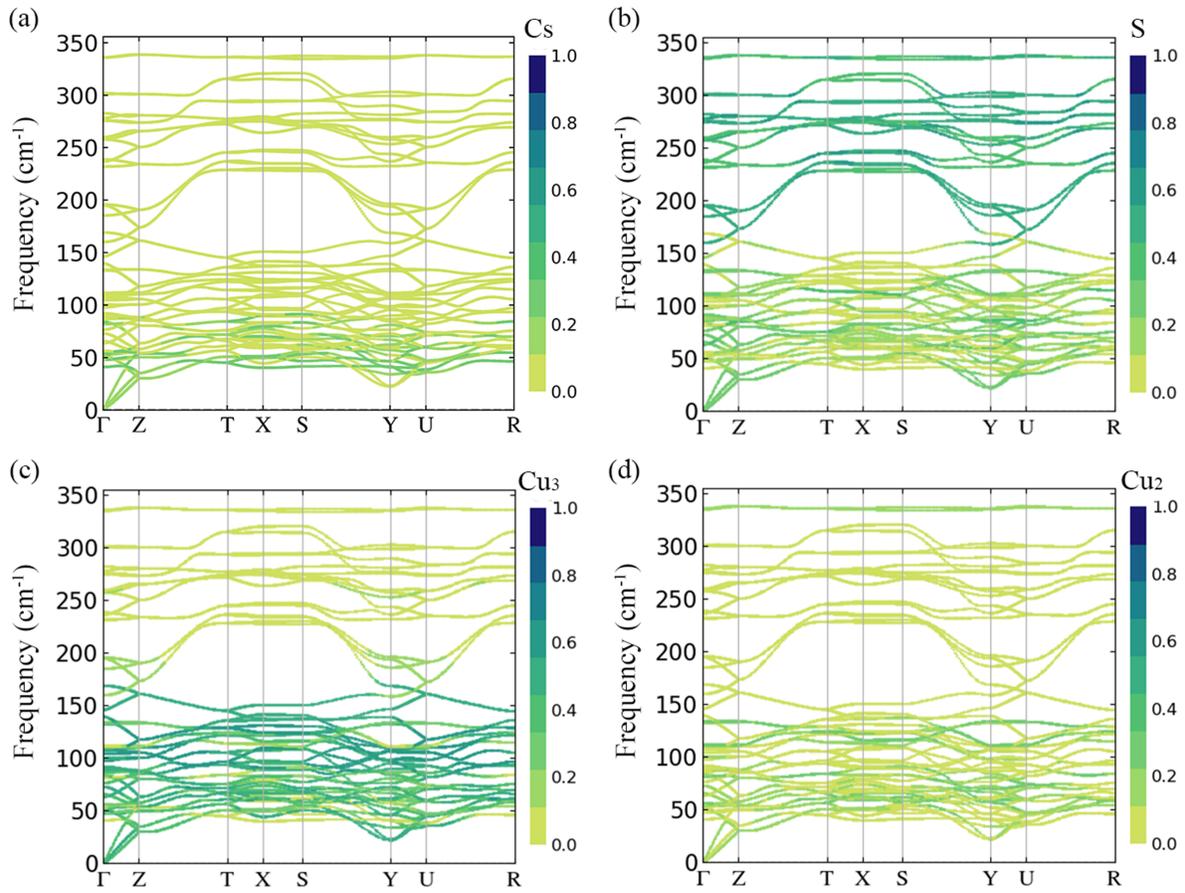}
  \caption{(a) The color-coded atomic participation ratio (APR) of Cs atoms projected onto the phonon dispersions at 300 K. (b) The same as (a) but for S atoms. (c) The same as (a) but for Cu$_3$ atoms. (d) The same as (a) but for Cu$_2$ atoms.}
  \label{Fig3}
\end{figure}
\clearpage
\begin{figure}[t]
\centering
  \includegraphics[height=13cm]{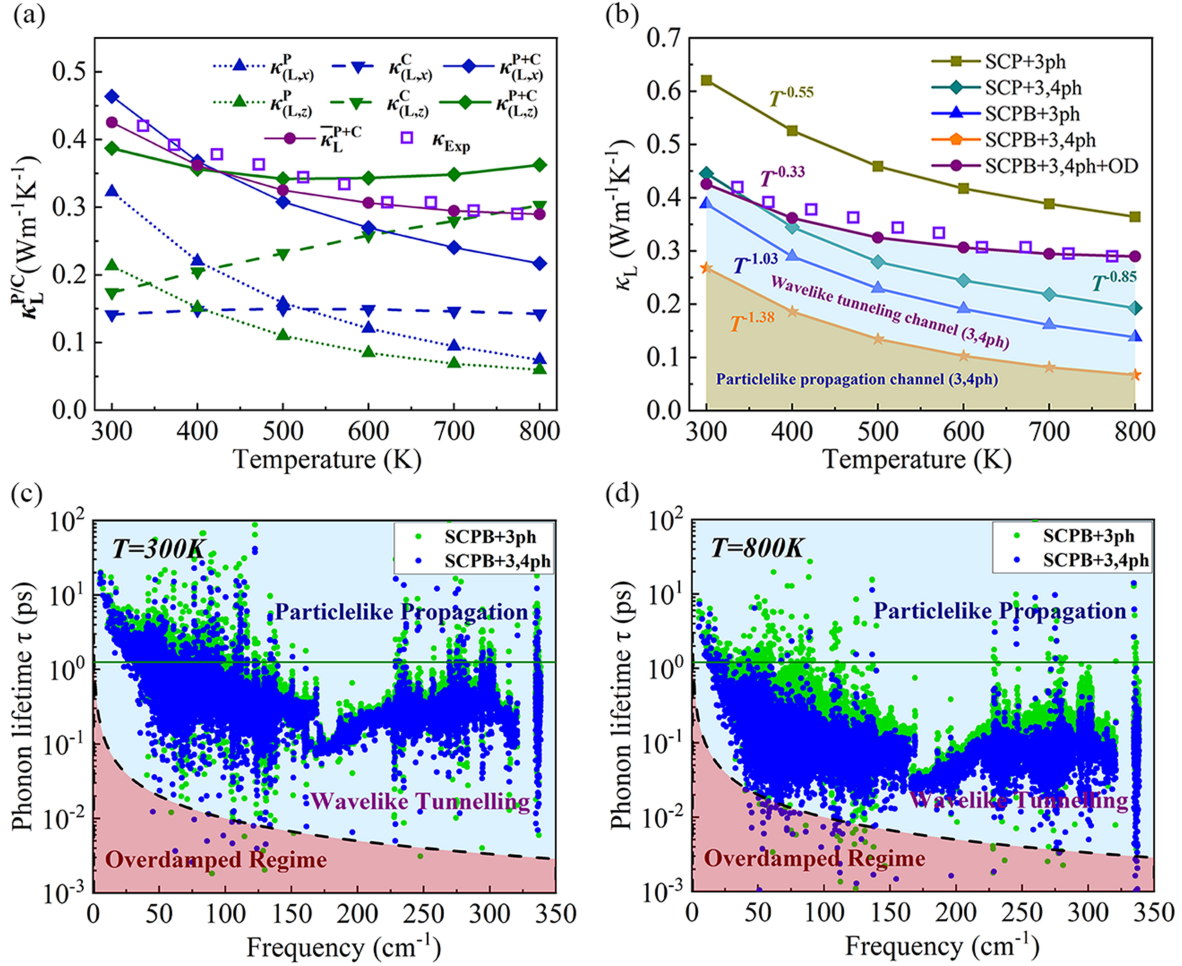}
  \caption{(a) Variations of temperature-dependent thermal conductivity including the contributions from particle-like ($\kappa^\mathrm{P}_\mathrm{L}$) and wavelike ($\kappa^\mathrm{C}_\mathrm{L}$) conduction mechanisms by using the SCPB+3,4ph+OD model. 
  (b) Temperature dependence of $\kappa_\mathrm{L}$ based on various theories including the SCP/SCPB + 3/4ph, SCPB + 3,4ph + OD models. The SCPB+3,4ph+OD model reveals that the phonon and coherence channels contribute to thermal transport in the orange and blue regions, respectively. (c) Calculated phonon lifetime as a function of frequency at 300 K, where black dotted line and green solid line represent the Ioffe-Regel limit and  Wigner limit in time. (d) The same as (c), but for 800 K.}
  \label{Fig4}
\end{figure}
\clearpage
\begin{figure}[t]
\centering
  \includegraphics[height=12.8cm]{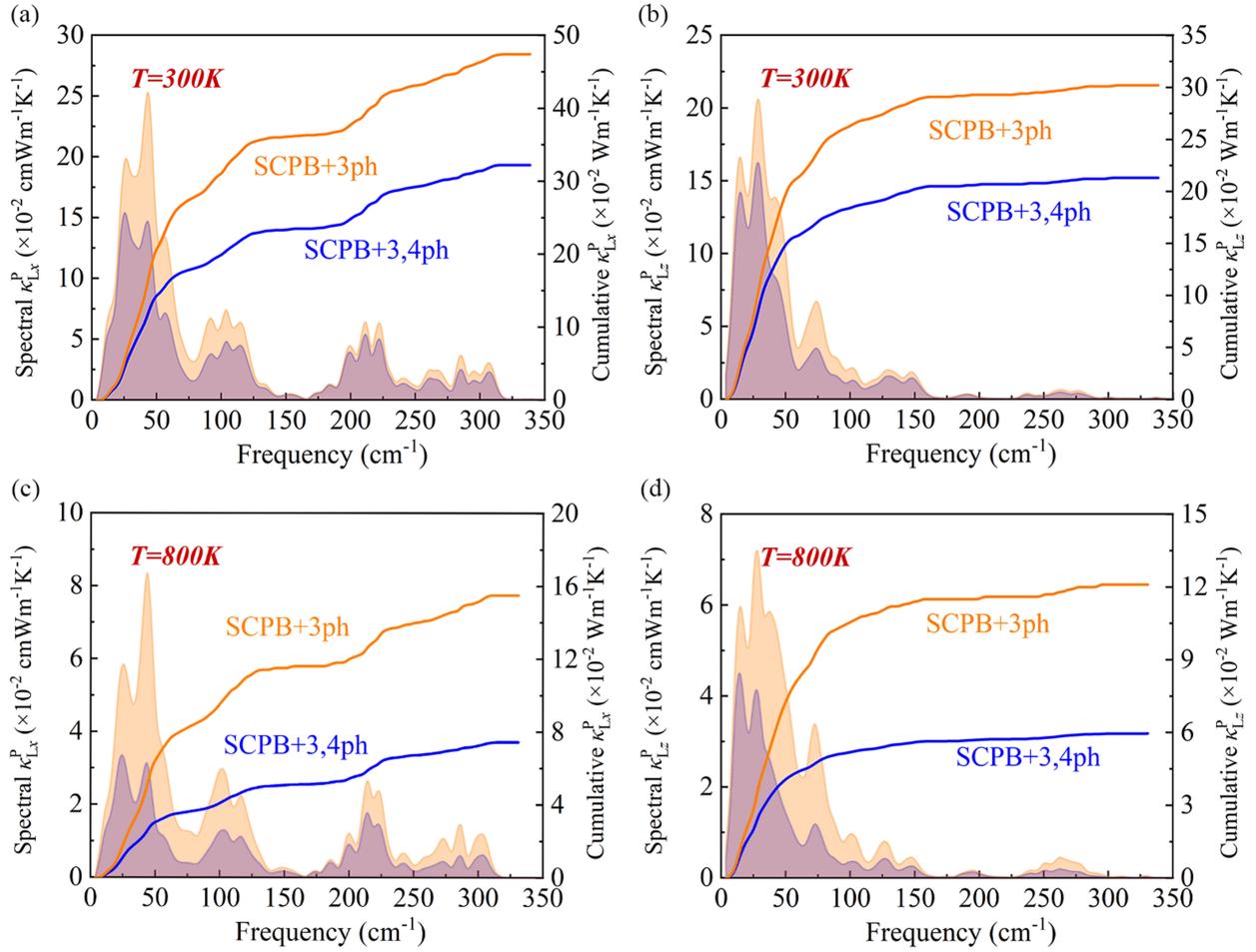}
  \caption{Calculated spectral/cumulative populations’ thermal conductivity using the SCPB+3ph and SCPB+3,4ph models, at 300 and 800 K, respectively. (a) Spectral and cumulative populations’ thermal conductivity along \textit{x} axis at 300 K. (b) The same as (a) but along \textit{z} axis. (c) The same as (a) but for 800 K. (d) The same as (b) but for 800 K.}
  \label{Fig5}
\end{figure}
\clearpage
\begin{figure}[t]
\centering
  \includegraphics[height=17.5cm]{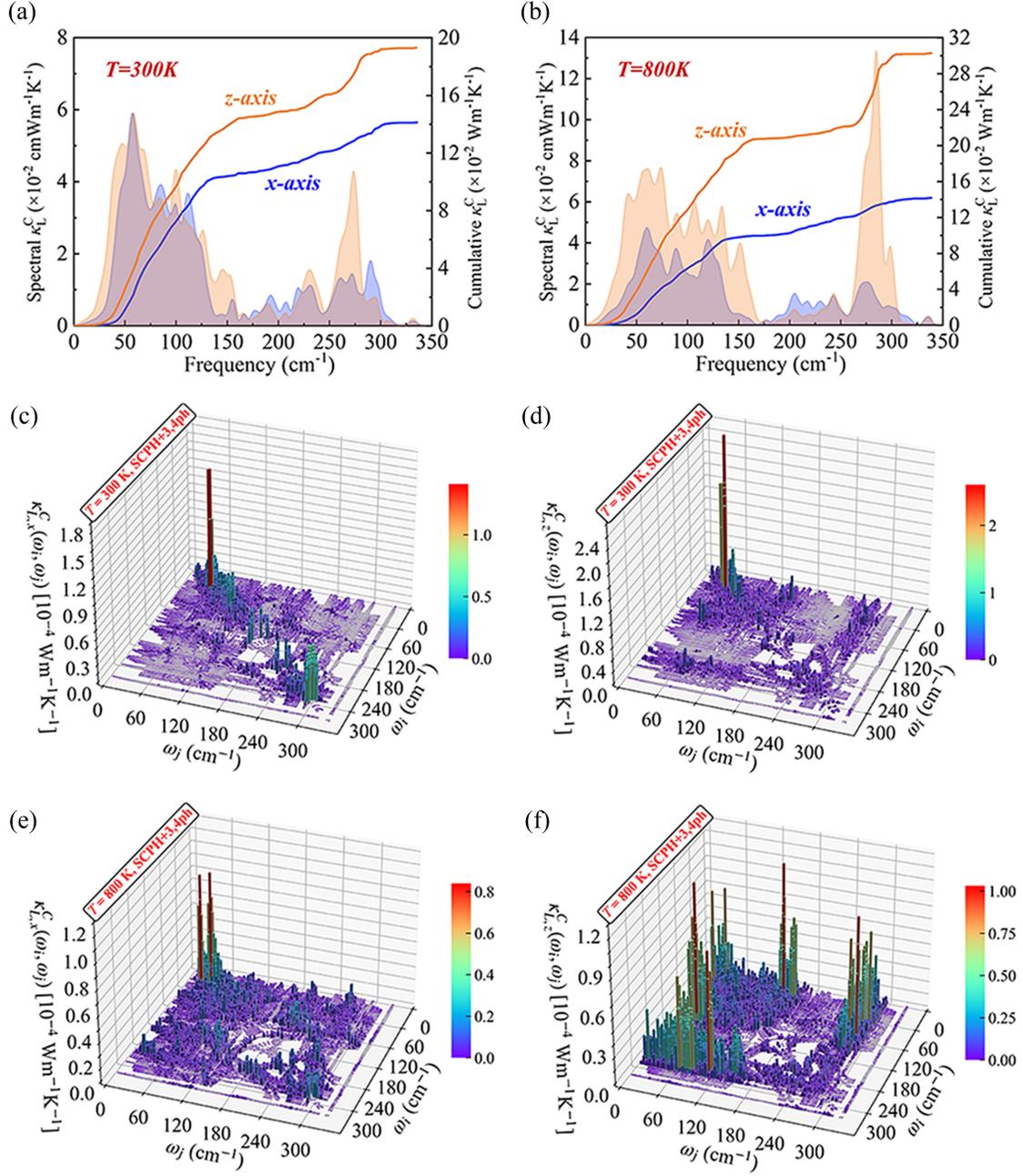}
  \caption{(a) Spectral and cumulative coherences’ thermal conductivity along \textit{x} and \textit{z} direction at 300 K. (b) The same as (e), but for 800 K. (c) Three-dimensional visualizations modal $\kappa^\mathrm{C}_\mathrm{L}$($\omega_{qj}$,$\omega_{qj'}$) of the contributions to the coherences' thermal conductivity calculated by the SCPB+3,4ph+OD model along with the \textit{x} axis at 300 K. The diagonal data points ($\omega_{qj}$ = $\omega_{qj'}$) indicate phonon degenerate eigenstates. (d) The same as (c), but for \textit{z} axis. (e) The same as (c), but for 800 K. (f) The same as (d), but for 800 K.}
  \label{Fig6}
\end{figure}
\clearpage
\begin{figure}[t]
\centering
  \includegraphics[height=12cm]{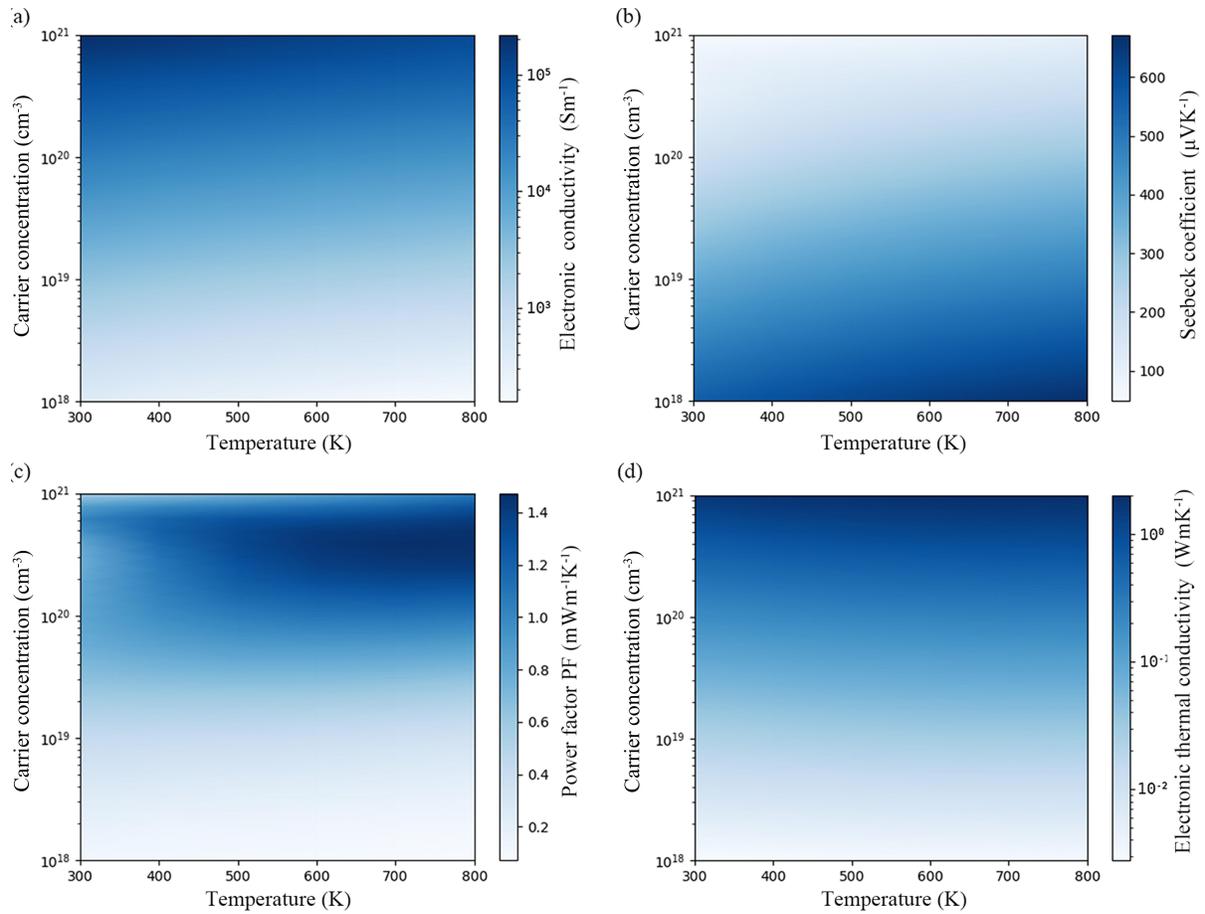}
  \caption{Calculated \textit{p}-type electronic transport properties involving (a) electrical conductivity $\sigma$,  (b) Seebeck coefficient $S$, (c) power factor PF, (d) electrical thermal conductivity $\kappa_e$, corresponding to carrier concentrations from 10$^{18}$ cm$^{-3}$ to 10$^{21}$ cm$^{-3}$ and temperatures from 300 K to 800 K.}
  \label{Fig7}
\end{figure}
\clearpage
\begin{figure}[t]
\centering
  \includegraphics[height=6.3cm]{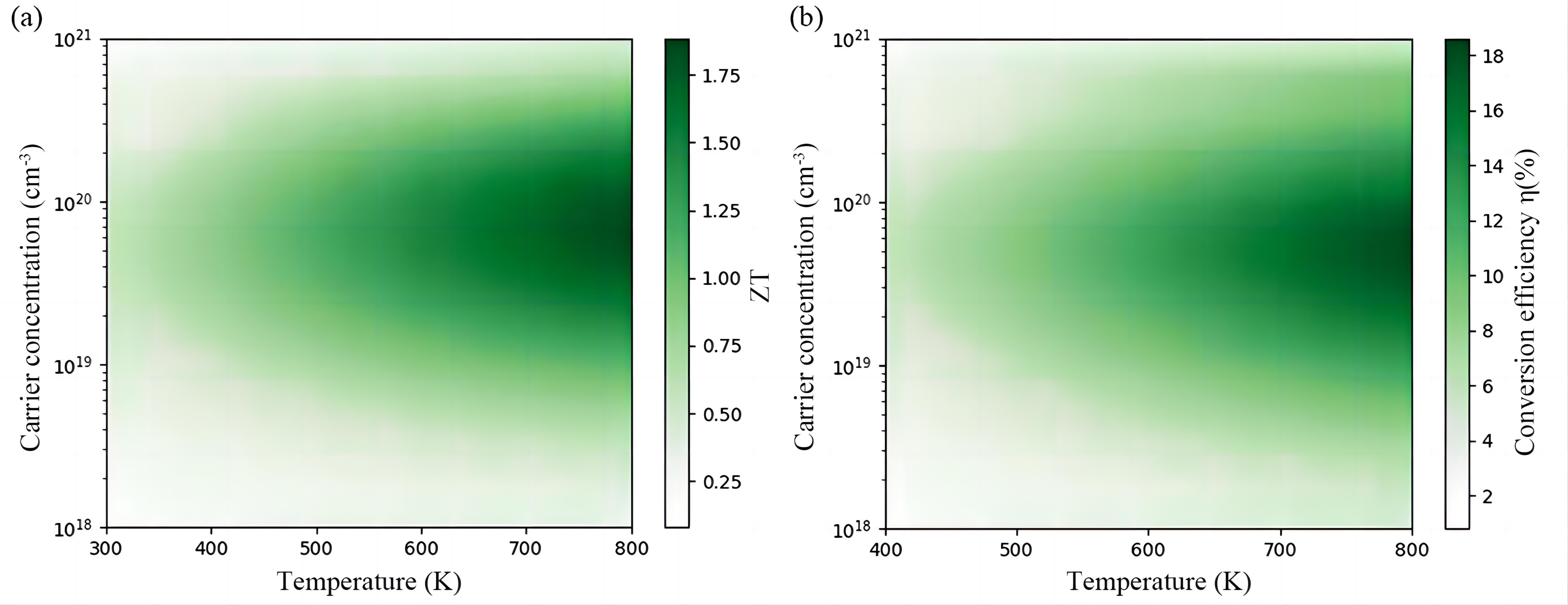}
  \caption{(a) Calculated p-type thermoelectric figure of merit ZT corresponding to carrier concentrations ranging from 10$^{18}$ cm$^{-3}$ to 10$^{21}$ cm$^{-3}$ and temperatures spanning from 300 K to 800 K. (b) Theoretically optimal thermoelectric conversion efficiency $\eta$ (\%) corresponding to carrier concentrations from 10$^{18}$ cm$^{-3}$ to 10$^{21}$ cm$^{-3}$ and hot end temperatures from 400 K to 800 K, where the cold end defaults to 300 K.}
  \label{Fig8}
\end{figure}
\end{document}